\documentstyle[prl,aps,psfig]{revtex}
\begin{document}
\twocolumn[\hsize\textwidth\columnwidth\hsize\csname
@twocolumnfalse\endcsname

\tighten
\draft

\title{Limits on Neutrino Mass from Cosmic Structure Formation}
\author{Masataka Fukugita$^{1,2}$, Guo-Chin Liu$^3$, Naoshi Sugiyama$^3$}
\address{$^1$ Institute for Cosmic Ray Research, University of Tokyo,
Tanashi, Tokyo 188, Japan}
\address{$^2$ Institute for Advanced Study, Princeton, NJ 08540, U. S. A.}
\address{$^3$ Department of Physics, Kyoto University, Kyoto 606, Japan}

 
\date{\today}

\maketitle

\begin{abstract}
We consider the effect of three species of neutrinos with nearly
degenerate mass on the
cosmic structure formation in a low matter-density universe
within a hierarchical clustering scenario
with a flat initial perturbation spectrum. 
The matching condition for
fluctuation powers at the COBE scale and at the cluster scale
leads to a strong upper limit on neutrino mass. For 
a flat universe with matter density parameter $\Omega=0.3$, we obtain 
$m_\nu<0.6$ eV for the Hubble constant $H_0<80$ km s$^{-1}$ 
Mpc$^{-1}$. Allowing for the more generous 
parameter space limited by $\Omega<0.4$, $H_0<80$  
km s$^{-1}$ Mpc$^{-1}$ and age $t_0>11.5$ Gyr, the limit is 0.9 eV.

\end{abstract}


]

Recent experiments for atmospheric and solar neutrino fluxes suggest
that the neutrinos are massive. In particular, the atmospheric neutrino
experiment indicates an almost maximal mixing between the two neutrinos, which
is most naturally understood if the relevant 
species  are nearly degenerate
in mass.  Nearly maximal mixing is also a viable possibility to explain
the long-standing solar neutrino problem with oscillation either in vacuum or
in matter, although there remains the 
solution that it is explained by small-angle mixing via 
oscillation in matter [1].
For these reasons the idea has gained popularity that the three 
neutrinos are massive and almost degenerate in mass (e.g., [1,2]).
The degenerate neutrinos mean that neutrino mass is larger than 
several tenths of eV, and  this means that they
provide the universe with a matter density comparable to or more than
that in stars, and play some role in cosmological structure formation.

There are a few authors who discussed the possibility that neutrinos have
played an active role in the formation of large-scale structure of
the universe, especially in giving a power at a large scale 
which otherwise cannot be accounted for in the standard cold dark 
matter scenario at the critical matter density [3]. At the time of
the emergence of this idea theorists took more seriously the 
Einstein-de Sitter (EdS) universe
of the critical matter density, 
so that typical compositions of the 
matter were assumed to be $\Omega_{CDM}=0.7-0.8$ and $\Omega_\nu=0.3-0.2$ 
in units of the closure density, $10.54h^2$keV (cm)$^{-3}$, where 
 $h$ is the Hubble constant $H_0$
in units of 100 km s$^{-1}$ Mpc$^{-1}$.
This neutrino mass density corresponds to neutrino mass 
of $(30-20)h^2$ eV.  There have been many explorations
of this scenario since the proposal [4], and the current conclusion is that
the neutrino density in excess of $\Omega_\nu\geq0.3$ is disfavoured
in the EdS universe 
 from the viewpoint of early cosmic structure formation.

Over the last few years the evidence has been accumulated 
indicating a low density universe. There are also observations pointing
to the dominance of the vacuum energy (cosmological constant, $\Lambda$) 
that makes
universe's curvature flat, which is also preferred from theoretical
point of view for a low matter density universe.
The list in favour of a low matter-density universe 
includes: (1) Hubble constant - cosmic age mismatch for the $\Omega=1$
universe; (2) No positive indications for the presence of copious
matter beyond
the cluster scale: the mass to light ratio inferred 
 from clusters and galaxies, $M/L=(100-400)h$, corresponds to 
$\Omega=0.1-0.3$ [5]; (3) consistency of the cluster baryon fraction
with the field value [6];
(4) The slow evolution of the cluster abundance from redshift z=0 
to 0.8 together with the abundance normalization at z=0 [7]; (5) The
Hubble diagram of Type Ia supernovae [8]. The data indicate a finite
cosmological constant. In view of systematic errors 
in various steps of the analysis, however, a zero $\Lambda$ is probably 
not excluded, whereas an $\Omega=1$ universe is too 
far away from the observations; 
(6) The perturbation spectral-shape parameter $\Gamma=\Omega h=0.2-0.3$ from
large-scale structure [9]; (7) Matching of the power spectrum between COBE
and galaxy clustering [10];
(8) Evolution of small scale non-linear galaxy clustering [11]; (9) Local
velocity field versus density enhancement [12]. The results 
of (1)$-$(9) converge
between $\Omega_0=0.2$ and 0.4.
 
The two positive indications for the presence of a cosmological constant
are the Type Ia supernova Hubble diagram mentioned above and the  
acoustic peak distribution in the
cosmic microwave background radiation (CBR) anisotropies [13], which
says that the universe is close to flat whether dominated by matter or
vacuum energy.

We remark that the results from large-scale velocity flow analyses are 
controversial; the resulting $\Omega$ 
varies from analyses to analysis compounded by the uncertainty in the biasing
parameter regarding the extent to which 
galaxies trace the mass distribution [14].  
Recent analyses 
of galaxy peculiar velocities combined with other observations
claim that they are consistent with 
a low density universe with 
a finite $\Lambda$ [15].

There seems no extensive analysis available for the effect of neutrinos in
a low matter-density universe, although a brief reference has been made in
[4].
In this paper we consider the effect of massive, {\it three~degenerate} 
neutrinos on the
cosmic structure formation in the low matter-density 
universe. 
We assume the hierarchical structure
formation dominated by cold dark matter
(CDM), the current standard model of the cosmic 
structure formation. The match of the power spectrum
in the COBE scale with that in the galaxy clustering
provides the most conspicuous evidence for the model.

While power spectrum analyses are the most common
way to demonstrate the consistency of the hierarchical clustering
scenario, the amplitude estimated from galaxy clustering receives
unknown biasing factors associated with galaxy formation mechanisms;
therefore, this is not appropriate for a quantitative analysis
as presented here. We consider matching of the two normalizations 
of the cosmic mass density fluctuation power, 
the normalizations derived from COBE 
at a several hundred Mpc scale
and the rich cluster abundance at z=0 which measures the power
at $\approx8h^{-1}$ Mpc scale and most conveniently
represented by the rms mass fluctuation parameter $\sigma_8$. 
We do not use the $\sigma_8$ parameters derived from velocity fields
or other observations, which are more susceptible to various
uncertainties.
The advantage of using the cluster abundance information 
is that it refers to
the mass function that is not affected by any biasing uncertainties,
and the fiducial length scale 8 $h^{-1}$ Mpc is close to that of
clusters before collapse. 
The requirement of this matching leads us to derive quantitative 
constraints on 
the neutrino contribution to cosmic structure formation.

We first consider the flat universe with low matter density, 
but also discuss later the case for 
open universes. 
We assume a flat (Harrison-Zeldovich) initial power spectrum,
$P(k)=Ak^n $
with $n=1$, which is the most natural prediction of inflation. 
The fluctuation spectrum receives a modification as $Ak^nT(k)$ for
a large $k$ as the fluctuations evolve [16]. The shape of the 
transfer function $T(k)$ depends on the assumed cosmological model, and
the neutrino content.
We use the computer code CMBFast [17] to calculate the transfer function 
for many choices of parameters. The spectrum is normalized with a 
fitting formula around $\ell=10$ deduced by Bunn \& White [18]
 from the four-year COBE-DMR data [19]. They have estimated one standard
deviation error to be 7\% in square root of the harmonic coefficient of CBR
anisotropies $C_\ell$.
We take the baryon fraction
$\Omega_B=0.015h^2$ corresponding to $\eta_{10}=4$ [20].
The result is not very sensitive to this choice.

We calculate the specific mass fluctuations within a sphere of a radius
of 8 $h^{-1}$ Mpc by integrating the spectrum with a top hat 
window:

\begin{eqnarray}
\sigma_8^2&=&\langle (\delta M/M)^2\rangle_{r<8h^{-1}} \nonumber\\
&=&\int_{r<8h^{-1}}dk 4\pi k^2 |\delta_k|^2 \Big[3 {\sin(kr)-(kr)\cos(kr)
\over (kr)^3} \Big]^2 .
\end{eqnarray}
The resulting $\sigma_8$ for a given Hubble constant and the
neutrino mass density $\Omega_\nu$ is presented in
Fig. 1 for a flat universe (a) $\Omega=0.3$ and $\lambda=0.7$,
and (b) $\Omega=0.4$ and $\lambda=0.6$. A set of curves 
(increasing towards the right) gives
contours of constant $\sigma_8$. Another set of curves shows the 
neutrino mass density for given neutrino mass. 

The neutrino mass that concerns us is in the range $<1 $eV for most  
cases. The core radius of neutrino clustering
allowed from the phase space argument [21] is
\begin{equation}
R_\nu=3.2{ \rm Mpc}(m_\nu/1 {\rm eV})^{-2}(v/1000 {\rm km~s}^{-1})^{1/2} , 
\end{equation}
which is large compared with the core radius of rich clusters
$R_c\simeq(0.12\pm0.02)h^{-1}$Mpc$^{-1}$ [22] for velocity dispersion
$v\approx 10^3$ km s$^{-1}$. Together with small neutrino mass density,
we can ignore
the neutrino component in integrating the cluster mass. We 
have estimated the contribution from neutrinos to the cluster
mass within linear perturbation theory. The inclusion changes the
result at most by a few percent, which can safely be ignored in the
present argument.

\begin{figure}[htbp]
    \centerline{\psfig{figure=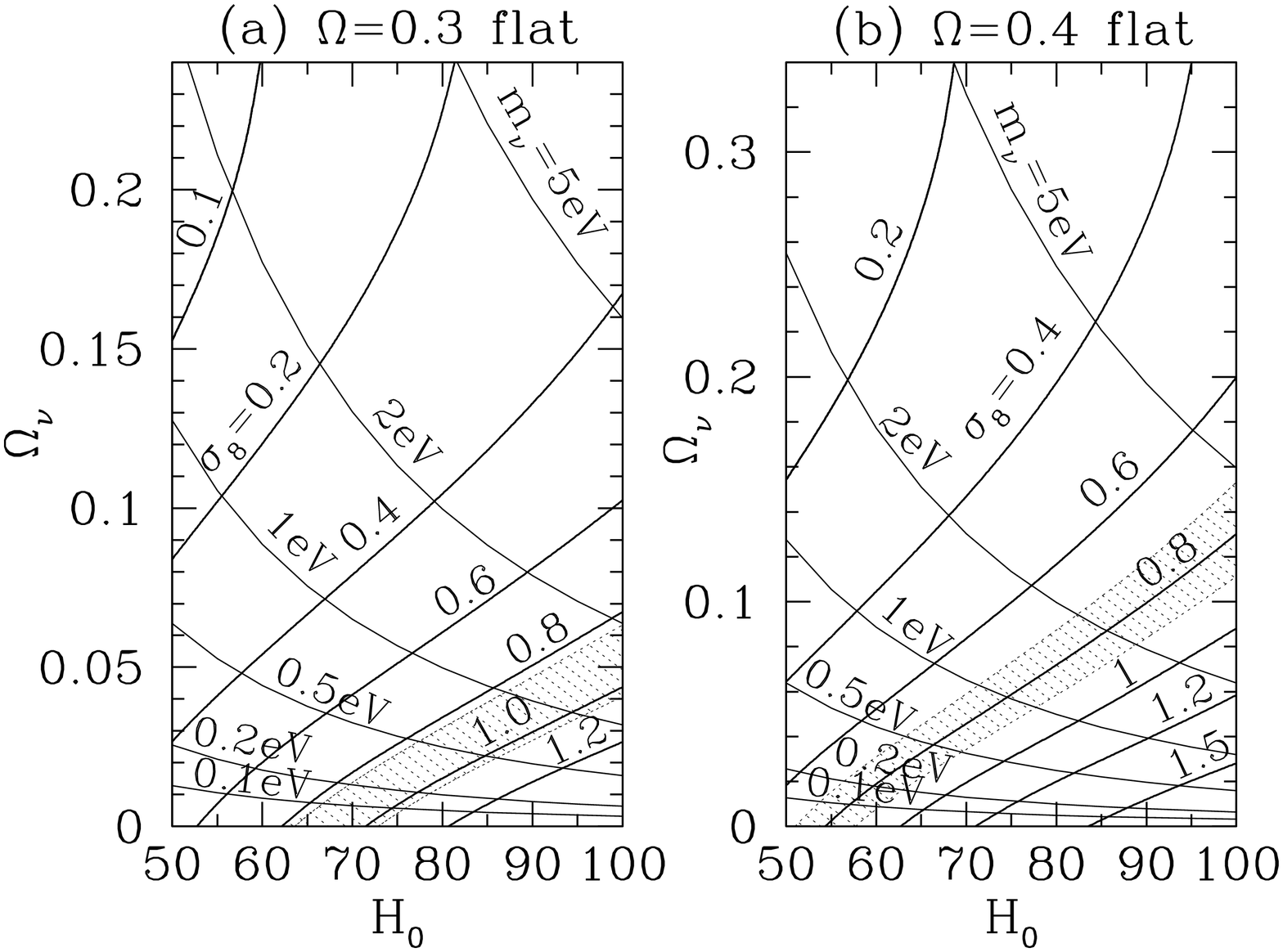,width=9.0cm}}
      \caption{
Contours of constant $\sigma_8$ derived from the COBE normalization
in the $H_0-\Omega_\nu$ plane for flat universes ($\Omega+\lambda=1$).
The region that satisfies the matching condition with the cluster 
abundance is indicated by a shade.  
Another set of curves indicates 
netrino mass density $\Omega_\nu=(3m_\nu/93.84{\rm eV})h^{-2}$.
(a) $\Omega=0.3$, (b) $\Omega=0.4$.
}      \label{fig:flat}
\end{figure}

This calculated $\sigma_8$ is compared with the value estimated from 
the rich cluster abundance. The estimate of $\sigma_8$ 
has been made by a number of authors [23-25,6]. 
The most ambiguous in such analyses
is the estimate of the cluster mass, but the modern results are 
well converged among the authors,
at least for $z\approx 0$ clusters.  A summary is presented in Table 1.
We take the values given by Eke et al. [24] which agree with
other estimates within the error:
$\sigma_8=0.93\pm0.07$ for $\Omega=0.3$ and $\sigma_8=0.80\pm0.06$ 
for $\Omega=0.4$. Adoption of Viana \& Liddle's [25] value
makes the derived limit
on neutrino mass slightly tighter. If we add the normalization error of the
CBR anisotropies in quadrature, the errors
become 0.10 and 0.09, respectively. The allowed range is shown by
shadows in the figure.

We see from Fig. 1 that one can obtain the limit on neutrino mass if
the Hubble constant is set. For $H_0=70$ km s$^{-1}$ Mpc$^{-1}$
we obtain 0.21 eV ($r=\Omega_\nu/\Omega\leq 5$\%)
for $\Omega=0.3$ and 0.91 eV ($r\leq 15$\%) for $\Omega=0.4$.
For $H_0=80$ km s$^{-1}$ Mpc$^{-1}$ our limits are 0.62 eV ($r\leq10$\%)
for $\Omega=0.3$ and 1.8 eV ($r\leq 22$\%) for $\Omega=0.4$.
Our limit is summarized by a fitting formula: 
\begin{equation}
m_\nu<[5.20h(\Omega/0.3)^{2.03}-3.20(\Omega/0.3)^{1.32}]h^2.
\end{equation}
Allowing for conservative parameter space,
$\Omega\leq0.4$, $H_0\leq80$ km s$^{-1}$ Mpc$^{-1}$
and $t_0>11.5$ Gyr, the upper limit is 0.87 eV, which
corresponds to $r\leq 13$\% of the total mass density.

\begin{figure}[htbp]
    \centerline{\psfig{figure=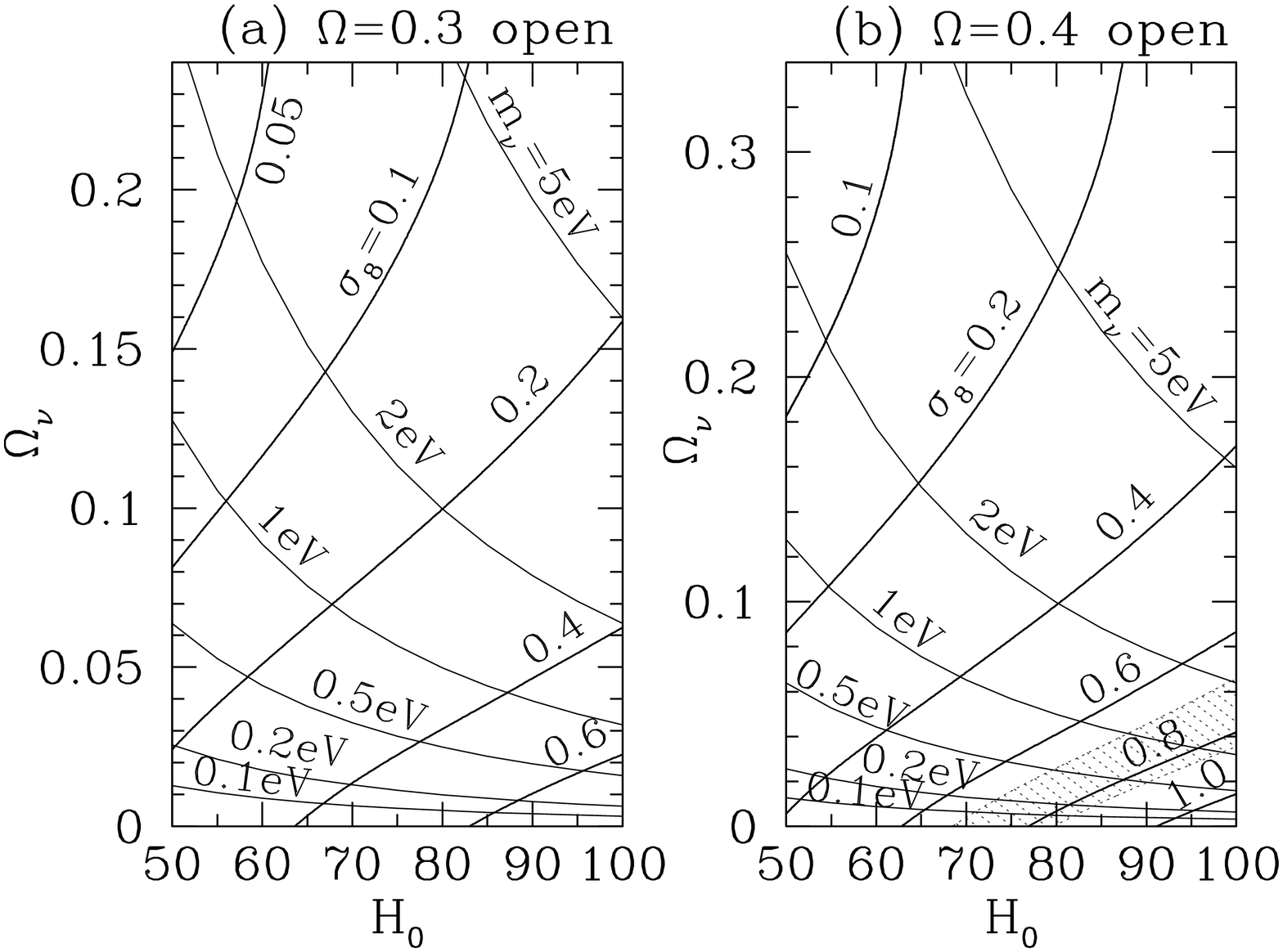,width=9.0cm}}
      \caption{
Same as Fig. 1, but for open universes.
}      \label{fig:open}
\end{figure}

A similar figure is given in Fig. 2 for zero-$\Lambda$ universes,
(a) for $\Omega=0.3$ and (b) for $\Omega=0.4$. The 
normalization from the cluster abundance is $\sigma_8=0.76\pm0.09$
and 0.87$\pm$0.09 including the CBR normalization error.
  There is no consistent parameter range
for $\Omega=0.3$ for $H_0<100$ km s$^{-1}$ Mpc$^{-1}$
with or without neutrinos. A consistent parameter range
appears for $\Omega=0.4$, but only with a relatively high 
$H_0$. No-neutrino models are consistent for $70<H_0<80$
km s$^{-1}$ Mpc$^{-1}$.
Requiring  $H_0\leq 80$ leads to $m_\nu<
0.5$ eV, which is significantly stronger than the one for 
the flat universe. 

The modification of power spectra with inclusion of massive neutrinos 
has been discussed by Hu et al. [26]. The change is about by a factor
of two for $m_\nu= 1$ eV when $H_0=70$ km s$^{-1}$
Mpc$^{-1}$.
The scatter among the various data and our ignorance of the biasing
factor make it difficult to exclude 1 eV neutrinos using the current
power spectrum data. 
A strong constraint, such as $m_\nu<0.4$ eV, would be obtained only when 
the power spectrum is derived from a
homogeneous galaxy sample with statistics as high as that would
be expected in the Sloan Digital Sky Survey [26]. This would provide us
with an alternative mean to set a limit on the neutrino mass, although
we must still assume the biasing factor being scale independent. 
The reason we obtained a strong limit in this paper 
is ascribed to the advantage
of using the cluster {\it mass} function, which is 
directly related to the mass fluctuation, as well as 
using the information spanning a very large
baseline in the length scale.

Let us discuss possible uncertainties or loop-holes of our argument.
We have ignored the contribution from gravitational wave perturbations
to CBR spectrum. Its inclusion only makes the limit on 
neutrino component more stringent. A possible loop-hole in our argument
is the possibility that the index of the power spectrum is significantly
larger than one.  The COBE data alone do not exclude an index in the
range $0.9<n<1.5$, but the range is reduced to $1< n < 1.2$ if
supplemented by other CBR data on small scales [12].
With an index, $n<1$, which can be easily realized with 
inflation models, the constraints become tighter. If $n>1$, 
the excess large-scale 
power generated by neutrino perturbations
are cancelled by the intrinsically small large-scale power 
and more massive neutrinos become viable. For $n=1.2$, the limit
for $\Omega=0.3$ and $h=0.7$ is loosened from 0.2 to 0.7 eV, and
for $h=0.8$, 0.6 to 1.4 eV. For our $(H_0,\Omega,t_0)$ range discussed
above the limit is 1.8 eV, still quite strong. We remark that 
we need some tricky tuning to give $n>1$ in inflation models [27].

The limit we derived in this paper is quite strong. It is 5-20 
times stronger than would be 
obtained from a straightforward mass density consideration 
$\leq 93.8 \Omega h^2$ eV.
Ellis and Lola [2] have recently developed an argument for 
neutrinos with a degenerate mass as large as 5 eV with interesting
physics. Such neutrinos, however, bring 
a large mismatch into the fluctuation power between the very 
large scale and the cluster scale, causing a disaster to 
currently accepted cosmic structure formation
models.

Let us finally compare our limits
with those obtained from experiments or other cosmological considerations. 
A direct limit on electron 
neutrino mass from tritium beta decay 
is $\leq$ 4.4eV (95\% CL)[28] allowing for some systematic effects 
that make the measured $m_\nu^2$ negative. Additional limits are available
if the neutrinos are of the Majorana type. The limit of lifetime for 
double beta decay of $^{76}$Ge has now increased to $5.7\times10^{25}$ 
year (90\% CL)[29], which leads to $0.2-1.5$ eV depending on the 
nuclear matrix 
element used. We also refer to a limit from cosmological baryon excess:
the condition for baryon asymmetry left-over leads to the Majorana
neutrino to be $\leq 1-2$ eV [30]. The limit obtained in this paper
does not depend on neutrino types.
If one would consider only one species of neutrinos being 
massive, the mass limit simply becomes weaker by about a factor of
three.

\vskip10 mm
\noindent
{\bf Acknowledgements}

We thank George Efstathiou and Craig Hogan for usuful comments on the
draft manuscript.
MF and NS are supported by Grants-in-Aid of the Ministry of Education.
MF thanks the Newton Institute and Institute of Astronomy in Cambridge 
and NS the Max Planck Institut f\"ur Astrophysik for
their hospitality while this work was completed.




\begin{table}
\caption{
Summary of $\sigma_8$ deduced from the rich cluster abundance
at zero redshift.}
\begin{tabular}{cccccc}
 & EdS & \multicolumn{2}{c}{flat} & \multicolumn{2}{c}{open} \\
 ref.  & $\Omega=1$  &   $\Omega=0.4$  &   $\Omega=0.3$  &   
  $\Omega=0.4$ &    $\Omega=0.3$  \\ \hline
{1.} &   $0.52\pm 0.04$ &   $0.80\pm0.06$ &  $0.93\pm0.07$ & 
  $0.76\pm0.06$ &  $0.87\pm0.07$  \\
{2.} &   0.56  &  0.86  &   0.99  &    0.77  &  0.84  \\
{3.} &  0.53  &  0.79  & -----  & ----- &  0.80 \\
\end{tabular}
Note 1.-Eke et al. [24], 2.-Viana and Liddle [25], 3.-Bahcall et al. [6] 
\end{table}


\end{document}